\documentclass[letterpaper,twocolumn,10pt]{article}
\usepackage{usenix2019_v3}

\usepackage{tikz}
\usepackage{amsmath}
\usepackage{amsfonts}

\usepackage{graphicx}
\usepackage{stfloats}
\usepackage{subcaption}
\usepackage{listings}
\usepackage{algorithm}
\usepackage{algorithmic}
\usepackage{booktabs} 
\usepackage{xspace}

\graphicspath{ {./figure/} }

\newcommand{\system}{RTCUDB\xspace}

\begin{document}

\date{}

\title{\system: Building Databases with RT Processors}

\author{
{\rm Xuri Shi}\\
Fudan University\\
23210240273@m.fudan.edu.cn
\and
{\rm Kai Zhang }\\
Fudan University\\
zhangk@fudan.edu.cn
\and
{\rm X. Sean Wang}\\
Fudan University\\
xywangCS@fudan.edu.cn
\and
{\rm Xiaodong Zhang}\\
Ohio State U.\\
zhang@cse.ohio-state.edu
\and
{\rm Rubao Lee}\\
Freelance\\
lee.rubao@ieee.org
} 

\maketitle

\begin{abstract}

A spectrum of new hardware has been studied to accelerate database systems in the past decade. Specifically, CUDA cores are known to benefit from the fast development of GPUs and make notable performance improvements. The state-of-the-art GPU-based implementation, i.e., Crystal, can achieve up to 61$\times$ higher performance than CPU-based implementations. However, experiments show that the approach has already saturated almost all GPU memory bandwidth, which means there is little room left for further performance improvements.

We introduce \system, the first query engine that leverages ray tracing (RT) cores in GPUs to accelerate database query processing. \system efficiently transforms the evaluation of a query into a ray-tracing job in a three-dimensional space. By dramatically reducing the amount of accessed data and optimizing the data access pattern with the ray tracing mechanism, the performance of \system is no longer limited by the memory bandwidth as in CUDA-based implementations. Experimental results show that \system outperforms the state-of-the-art GPU-based query engine by up to \(18.3\times\) while the memory bandwidth usage drops to only 36.7\% on average.
\end{abstract}

\section{Introduction}

\label{sec:intro}

With ever-increasing data volume from applications, modern processors have been intensively studied to enhance the performance of database engines. Representative processors that database systems have been designed for include GPU CUDA cores~\cite{Crystal,HeavyDB,BlazingSQL}, Tensor cores~\cite{TCUDB,TQP}, and FPGAs~\cite{fpga1, fpga2}. 
Specifically, GPUs were originally designed to accelerate computer graphics. After they were found useful in general-purpose computing for the massive number of cores, the programming model evolved from OpenGL to CUDA/OpenCL, and GPUs are used to accelerate a broad class of data processing tasks.
Crystal~\cite{Crystal} is the state-of-the-art implementation of GPU databases, which has made a notable advancement in utilizing CUDA cores. Experimental results show that Crystal is \(16\times\) faster than the GPU-based HeavyDB and \(61\times\) faster than the CPU-based MonetDB.
However, experiments also show that Crystal saturates more than 97\% GPU memory bandwidth for queries in Star Schema Benchmark (SSB)~\cite{SSB}. Since the approach has already tried to minimize the amount of data accessed, it has become hard to improve the query performance on CUDA cores further.

Recently, commodity GPUs have incorporated ray tracing (RT) cores to boost the real-time rendering of 3D scenes. RT cores efficiently trace rays through a 3D space to identify intersected objects. With user-defined functions (a.k.a., shaders), RT cores can perform customizable operations upon ray-object intersections, providing versatility for various tasks.
RT cores have been leveraged across various scenarios to expedite data processing like K-nearest neighbor search~\cite{RT-KNNS, RTNN}, scan operator~\cite{RTScan, RTIndex}, and range minimum queries~\cite{range-min-query}. As an important type of architecture, RT cores have been adopted in mobile, desktop, and workstation processors, which are under fast development. Moreover, existing work like RTScan~\cite{RTScan} already shows that RT cores can bring up to $4.6\times$ higher performance than CUDA cores and CPU. Therefore, we believe RT cores have demonstrated the potential to become another critical architecture for general-purpose data processing tasks. 

Unlike database implementations on CUDA cores, accelerating a data processing program with RT cores requires the program to be mapped to an efficient ray tracing job.
In a ray tracing job, data records are transformed into primitives such as triangles or spheres positioned in a three-dimensional space with a bounding volume hierarchy (BVH), while a query is converted into rays in a specified region. 
If the task does not fit such a job mapping, or the mapping is inefficient (e.g., lack of parallelism with a limited number of rays), it may result in even lower performance than CPUs and CUDA cores~\cite{RTScan, RTIndex}.
Due to the above reason, exploring RT cores to accelerate database queries is quite challenging because an operator like \texttt{Join} or \texttt{GroupBy} is hard to transform into an independent RT job.
Moreover, since the execution of an operator depends on the output of its previous operator in the query plan, the BVHs for the following operators have to be built during query execution, which is very time-consuming.
Therefore, simply implementing independent RT-based operators like CUDA-based databases cannot exploit the performance advantage of RT cores.

\begin{figure*}
    \centering
    \includegraphics[width=\textwidth]{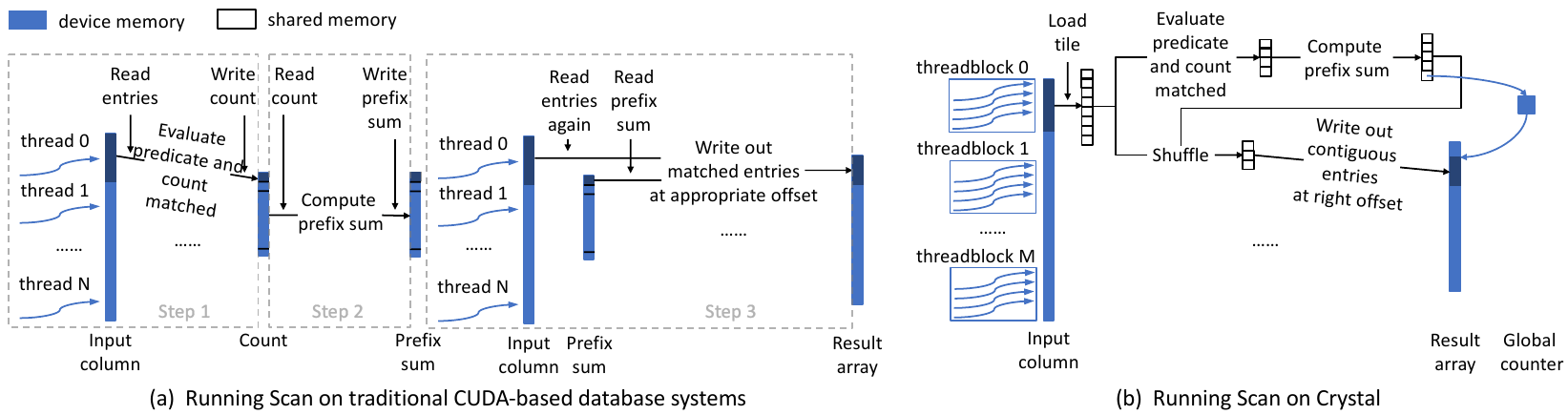} 
    \caption{Running Scan on CUDA-based database systems}
    \label{fig:database_arch}
\end{figure*}

In this paper, we propose \system, a query engine that uses ray-tracing cores to achieve unprecedented query performance. 
Instead of implementing an RT job for each operator, the main idea of \system is to map the core query execution containing multiple operators into a single ray tracing job. To be specific, \system maps the data attributes involved in \texttt{Aggregation}, \texttt{GroupBy}, and \texttt{Scan} to the coordinates \textit{x}, \textit{y}, \textit{z} of a data record in the 3D space, respectively. When building a BVH for the data records, the attributes involved are extracted and encoded as the primitive coordinates. This step can be viewed as performing the \texttt{Join} operation, and the BVHs can be taken as materialized views in the databases.
With the BVH for a query, rays in \system are launched in the region specified by \texttt{Scan} so that they intersect only with primitives that satisfy the predicates, significantly reducing the amount of data accessed. For each primitive, the data attributes involved in \texttt{Aggregation} and \texttt{GroupBy} are encoded in their 3D coordinates, which can be accessed directly for the corresponding processing. Therefore, another main advantage for \system is that, for each data record, \system can retrieve all the data attributes needed for the three operators with only one memory access, i.e., the access to the coordinate of a primitive, dramatically reducing random memory accesses. Moreover, the entire process is accelerated by ray tracing cores, which are designed for performing tasks like this. After RT processing, \system launches CUDA cores to execute other operators like \texttt{Having} and \texttt{OrderBy} in the query. 


The contributions of this paper are as follows.
\begin{itemize}
    \item We propose \system, a query engine that leverages RT cores for acceleration by effectively mapping a database query to a ray tracing job.
    \item We propose a set of encoding and ray launching mechanisms for operators so that \system can support complex queries with multiple attributes.
    \item We implement the prototype of \system and evaluate its performance under diverse workloads and configurations on the latest GPU.
\end{itemize}

Experimental results show that \system can significantly enhance the query performance. Compared with the state-of-the-art CUDA-based method, \system can improve the query performance by up to 18.3$\times$. At the same time, the usage of memory bandwidth drops from 97.4\% to only 36.7\% on average, which proves that \system has broken through the limitation of memory bandwidth. To our knowledge, it is the first work that demonstrates RT cores can be effectively used in building a database engine with unprecedented performance.

\section{Background and Motivation}

\label{sec:bam}

\subsection{Characteristics of Crystal}
\label{sec:bam:crystal}
\begin{figure*}[htbp]
    \centering
    \begin{minipage}[t]{0.35\textwidth}
        \centering
        \includegraphics[width=\textwidth]{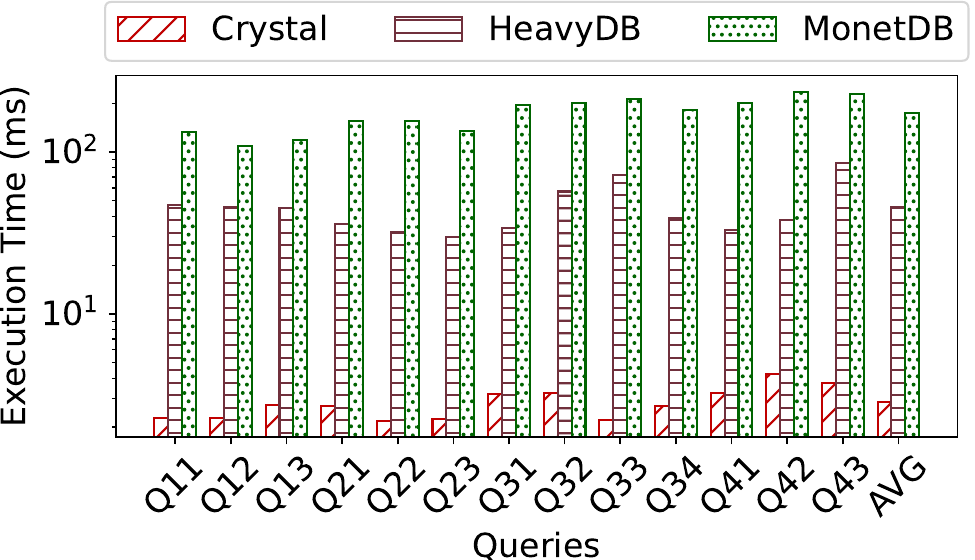}
        \caption{Comparison of query runtime for Crystal, MonetDB, and HeavyDB on SSB flat}
        \label{fig:MHC_ssb_fig}
    \end{minipage}
    \hfill
    \begin{minipage}[t]{0.35\textwidth}
        \centering
        \includegraphics[width=\textwidth]{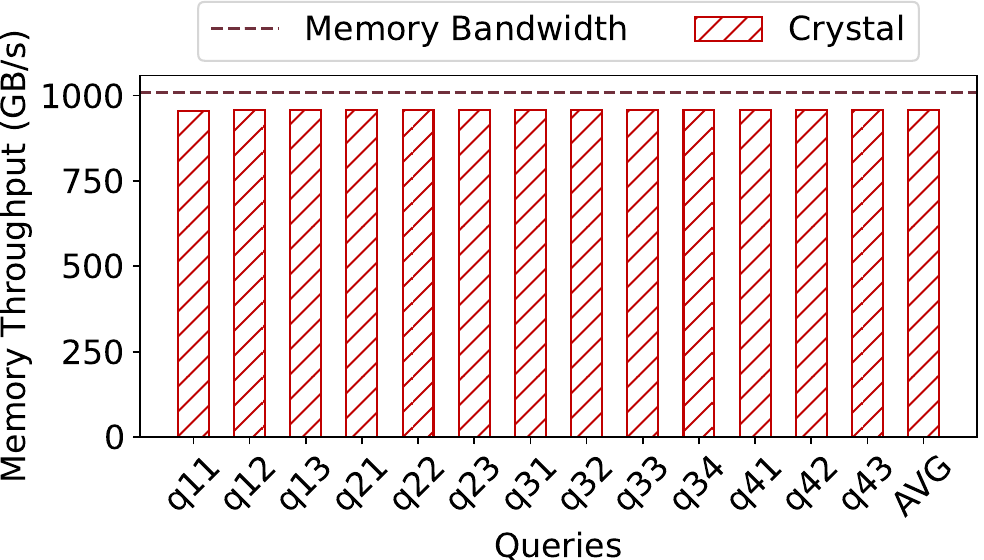}
        \caption{Query memory throughput of Crystal on SSB flat}
        \label{fig:bandwidth_simple}
    \end{minipage}
    \hfill
    \begin{minipage}[t]{0.28\textwidth}
        \centering
        \includegraphics[width=\textwidth]{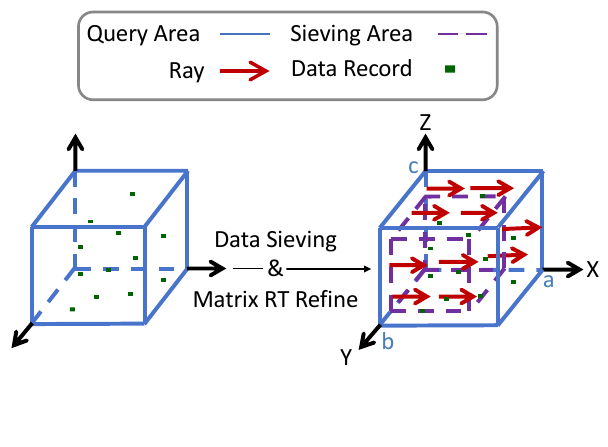}
        \caption{The execution of RTScan}
        \label{fig:rtscan}
    \end{minipage}
\end{figure*}

As a general-purpose computing device, GPU CUDA cores have been intensively studied to build high-performance query engines, such as Crystal~\cite{Crystal}, HeavyDB~\cite{HeavyDB}, BlazingSQL~\cite{BlazingSQL}, etc. Crystal is a state-of-the-art GPU database system based on CUDA cores. Figure~\ref{fig:MHC_ssb_fig} compares the query runtime of Crystal with HeavyDB and MonetDB~\cite{MonetDB} on SSB flat. The GPU used in the comparison is NVIDIA GeForce RTX 4090, and the CPU used is Intel(R) Xeon(R) Gold 5318Y. The results show that Crystal is, on average, \(16\times\) faster than HeavyDB and \(61\times\) faster than MonetDB. 

The key to Crystal's high performance is that its \textit{tile-based execution model} aims at efficiently utilizing the GPU shared memory, which has an order of magnitude higher bandwidth than the device memory (36618 GB/sec versus 1008 GB/sec in RTX 4090). 
Taking the \texttt{Scan} operator as an example, the traditional CUDA-based database systems execute the operator in three steps, as shown in Figure~\ref{fig:database_arch}a. In the first step, numerous CUDA threads are launched, and each thread evaluates the assigned entries and counts the number of matched ones. The count result is written to the \textit{count} array in the device memory, where the \textit{i}th element records the number of entries matched by thread \textit{i}. The second step computes the prefix sum of the \textit{count} array and stores the results in the \textit{prefix sum} array, which indicates the offset at which each thread should write.
In the last step, the same amount of threads as in the first step are launched. Each thread reads the allocated entries again and writes the matched entries to the \textit{result} array at the position based on the \textit{prefix sum} array. There are three main performance issues with this approach, including 1) reading the input column twice from the device memory, 2) reading and writing intermediate structures like \(count\) array and \textit{prefix sum} array in the device memory, 3) each thread writes to a different location in the \(result\) array, resulting in numerous random writes on the device memory.

Crystal treats a block of threads (i.e., thread block) as the basic execution unit, and each thread block processes a tile of entries at a time. 
Figure~\ref{fig:database_arch}b demonstrates how Crystal works.
First, each thread block loads a tile of entries from the device memory into the shared memory. The threads in a thread block then evaluate the predicate on the tile in parallel and store the number of matched entries for each thread in the \textit{count} array in the shared memory. The thread blocks compute the prefix sum of the \textit{count} array and increase a global counter atomically to calculate the offset of each thread block. The matched entries are shuffled to form a contiguous array in the shared memory to be efficiently copied to the device memory at the right offset to avoid random writes. 

Overall, Crystal's approach minimizes the reads and writes to the device memory, reduces the materialization overhead, and avoids random writes, bringing significant performance advancement.
Figure~\ref{fig:bandwidth_simple} shows the memory throughput of Crystal on the 13 queries of SSB flat. It can be observed that its memory throughput is very close (an average of 97.4\%) to the GPU device memory bandwidth.
Since Crystal has tried to minimize the amount of device memory accesses by loading data into the shared memory, there is little room left for further optimization.

\subsection{Background of Ray Tracing}
\label{sec:bam:rtb}
Ray tracing is a rendering technique used in computer graphics to simulate the way rays interact with objects in a scene. It works by tracing the paths of rays as they travel through a 3D space. Objects in the three-dimensional space are represented as primitives, whose types include triangles, spheres, and even custom primitives. All primitives in the space are wrapped by bounding volumes, which are usually Axis-Aligned Bounding Boxes (AABBs). AABBs are then organized hierarchically as a tree known as the Bounding Volume Hierarchy (BVH). 
A ray tracing job utilizes a BVH to traverse the space and find intersected primitives with the rays.

Although BVH avoids a large number of potential ray-primitive intersection tests, the BVH traversal and intersection tests are still time-consuming. Since the Turing architecture~\cite{Turing}, NVIDIA GPUs are equipped with dedicated hardware, i.e., RT cores, to speed up BVH traversal. 
Besides the mainstream desktop and workstation GPUs like NVIDIA’s RTX 40 series
and AMD’s RX 7000, data center GPUs like NVIDIA A40 and T4 also support ray tracing.
Specifically, NVIDIA RTX 4090 GPU integrates 128 RT cores.
OptiX programming model~\cite{OptiX} is an application framework for building ray tracing jobs. In OptiX, each ray is mapped to a CUDA thread. CUDA threads generate rays with the specified ray origins and directions. Then, the control is transferred to RT cores, which accelerate BVH traversal and ray-triangle intersection tests. For NVIDIA GPUs, triangles are the built-in primitives, which allows RT cores to accelerate intersection tests when the primitives are triangles.
When other types of primitives are used, control is transferred to CUDA cores to perform intersection tests defined in Intersection Shader. 

\subsection{Expedite Data Processing with RT Cores}
RT cores have been utilized to accelerate various data processing like K-nearest neighbor search~\cite{RT-KNNS, RTNN} and range minimum queries~\cite{range-min-query}.
RTScan~\cite{RTScan} and RTIndex~\cite{RTIndex} are pioneering implementations that leverage RT cores to accelerate the \texttt{Scan} operator.
Specifically, RTScan~\cite{RTScan} achieves significant performance improvement by mapping the evaluation of entire conjunctive predicates into a ray tracing process. Experiments show that RTScan achieves up to \(4.6\times\) higher performance than BinDex, which is the state-of-the-art scan approach on CPUs. Figure~\ref{fig:rtscan} demonstrates the approach of RTScan when evaluating a query with three predicates. For each data record, the three data attributes involved in the predicates are used as the coordinates of the corresponding primitive. Assuming that the conjunctive predicates are \( (0 \leq x \leq a \wedge 0 \leq y \leq b \wedge 0 \leq z \leq c)\), then the query area is a cuboid with the origin as a vertex and three edges of length \(a\), \(b\), and \(c\) as shown in the figure. The data records satisfying the conjunctive predicates are all in the query area. To reduce the computation overheads on intersection tests, RTScan adopts Data Sieving, which uses pre-stored results to filter most data records (the dashed area). RTScan also proposes Matrix RT Refine to launch more rays to enhance the parallelism for intersections in the remaining region. These techniques add up to form the performance improvement over CUDA cores and CPUs.

After analyzing RTScan and other RT-based implementations, we summarize three key aspects of efficiently mapping a data processing task to a high-performance RT job.
1) \textit{Evaluating multiple operators in one job:} RTScan utilizes the 3D feature to evaluate three predicates simultaneously within one RT job, whose execution time is even lower than that of evaluating one predicate.
2) \textit{Reducing the amount of data accesses:} The evaluation of conjunctive predicates and the Data Sieving technique dramatically shrink the querying region, leading to a significantly smaller number of primitives for intersections.
3) \textit{Enhancing the parallelism:} RTScan segments a long ray into several small rays with spacing, which aims at maximizing the utilization of RT cores while balancing their load. 

\subsection{Challenges of Accelerating Query Processing with RT Cores}
\label{sec:bam:aeq}
For GPU acceleration with CUDA cores, operators are generally implemented as separate CUDA kernels and executed sequentially. RTScan demonstrates a standalone implementation of accelerating \texttt{Scan} with RT cores.
However, accelerating database queries with each operator implemented as a separate RT job faces several critical issues that are hard to address.

\noindent\textbf{Difficulties in the three-dimensional mapping of operators:} In order to achieve excellent performance, a database operator needs to be effectively converted into a ray tracing job, or the performance advantage of RT hardware cannot be effectively exploited. 
For instance, a naive scan implementation on RT cores can be $2.3\times 10^4$ times slower than the state-of-the-art CPU-based implementation~\cite{RTScan}.
However, the natures of some database operators make it hard to map the data in the 3D space and hard to convert operations into ray intersections. 
Therefore, except \texttt{Scan}, other operators like \texttt{Join} and \texttt{GroupBy} have not demonstrated superior performance on RT cores so far.

\begin{figure*}[htbp]
    \centering
    \begin{minipage}[t]{0.34\textwidth}
        \centering
        \includegraphics[width=\textwidth]{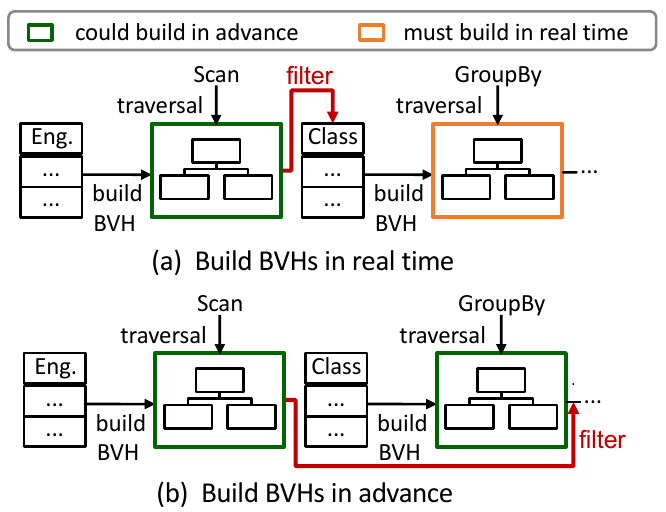} 
        \caption{Two strategies of query execution with RT cores}
        \label{fig:accelerateentirequery}
    \end{minipage}
    \hfill
    \begin{minipage}[t]{0.65\textwidth}
        \centering
        \includegraphics[width=\textwidth]{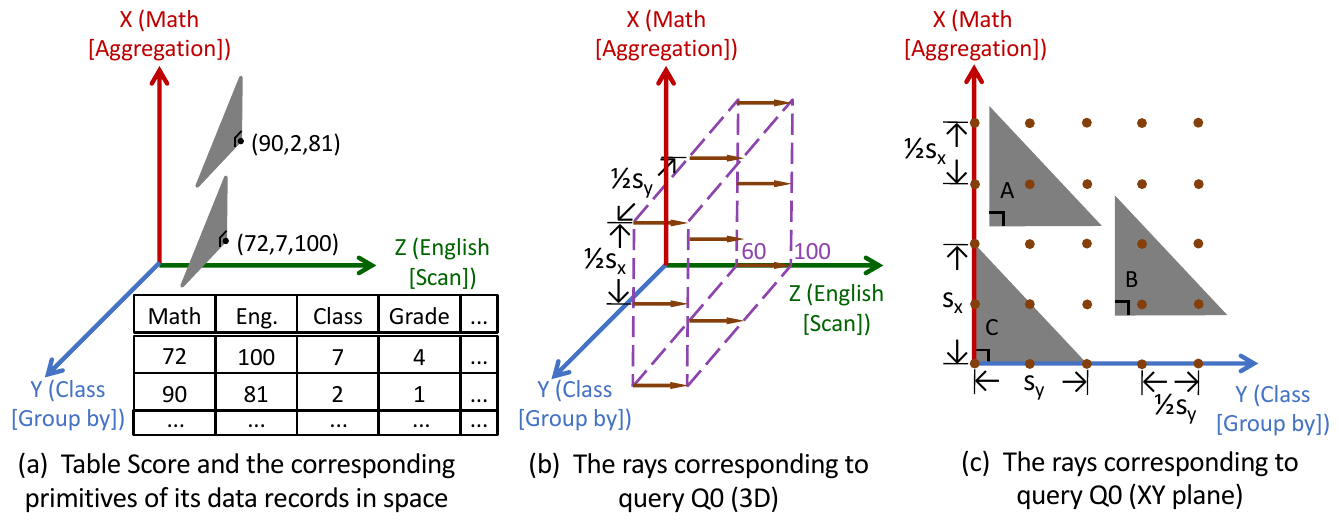} 
        \caption{The design of \system}
        \label{fig:query}
    \end{minipage}
\end{figure*}

\noindent\textbf{Inefficiency of the execution of multiple operators:} 
A ray tracing job uses a BVH as its index, but building a BVH is a time-consuming process that takes orders of magnitude longer time than the RT job itself. For a dataset with 120 million tuples, the average time to build a BVH is 227.84 ms, while the average time to launch rays is only 0.75 ms. When the operators of a query are executed sequentially on RT cores, the BVHs for the second to the last operator can only be built online, resulting in an ultra-high query processing latency. 

\begin{verbatim}
A: SELECT AVG(Math) 
   FROM Score 
   WHERE English >= 60 
   GROUP BY Class;
\end{verbatim}

We take the query \(A\) as an example, which executes the operators in the following order: \texttt{Scan} $\rightarrow$ \texttt{GroupBy} $\rightarrow$ \texttt{Aggregation}.
For \texttt{Scan}, the BVH it traverses can be pre-built from the English column directly. However, for \texttt{GroupBy}, the BVH it traversed is built from the Class column filtered by the execution results of \texttt{Scan}. Therefore, it can only start building the BVH after the execution of \texttt{Scan} is completed and the results are obtained. The above process is shown in Figure~\ref{fig:accelerateentirequery}a. As a result, the BVH building process for \texttt{GroupBy} has to be taken as a part of the query execution, which degrades the overall performance. 
An alternative scheme is shown in Figure~\ref{fig:accelerateentirequery}b, where the BVH for all operators have been pre-built from their corresponding attributes. In this case, \texttt{GroupBy} cannot get the filtered results from \texttt{Scan} and has to group all data on the column. Moreover, the results of \texttt{GroupBy} have to perform merge operations with the results from \texttt{Scan}. Consequently, this scheme completely negates the performance benefit of RT cores and is, therefore, severely inefficient. 

To conclude, due to the aforementioned issues, accelerating database queries with ray tracing cores to outperform CUDA cores and CPUs is particularly challenging.

\section{The Design of \system}

\label{sec:dai}

\subsection{Overview of \system}

\begin{figure}[bt]
        \centering
        \includegraphics[width=0.47\textwidth]{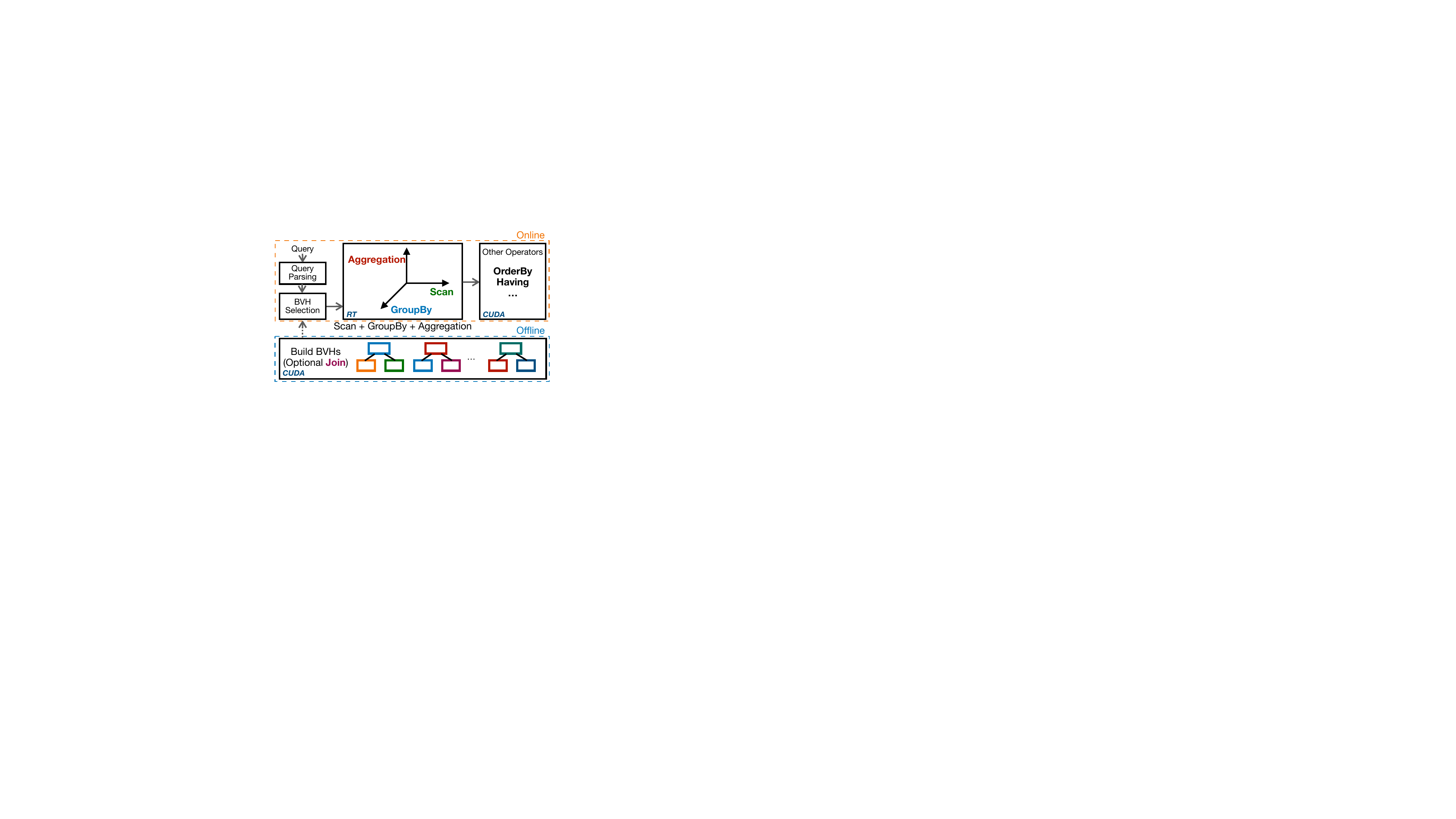}
        \caption{The workflow of \system}
        \label{fig:raydb_flow}
\end{figure}

We propose \system, a high-performance query engine accelerated by ray tracing cores. 
Instead of implementing each operator as a separate RT job, \system maps the execution of the core operators of a query, i.e.,  \texttt{Aggregation}, \texttt{GroupBy}, and \texttt{Scan}, to only one ray tracing job.
The basic idea is to use the attributes involved in the three operators as the coordinates for a data record, i.e., the attributes involved in \texttt{Scan} serve as the coordinate on the Z-axis, and the attributes in \texttt{Aggregation} and \texttt{GroupBy} are used as the X-coordinate and Y-coordinate, respectively.
To cope with queries where an operator involves multiple attributes, \system adopts different encoding schemes for each operator to map the attributes as one coordinate (Section~\ref{sec:ex}). Besides, encoding more attributes in each BVH can support more queries for execution.

The workflow of \system is illustrated in Figure~\ref{fig:raydb_flow}. First, BVHs are built offline on CUDA cores. The BVH building process needs to determine the coordinates of each primitive. Therefore, for data warehouses where the table is stored in a denormalized form, the attributes can be directly used. For databases with multiple tables, the involved tables need to be joined first. In this way, the BVHs can be viewed as materialized views, which is a common optimization in database systems. \system maintains multiple BVHs to handle queries involving different attributes. With a given query, \system parses the query and selects the BVH for it.

With the selected BVH, \system can map database queries into rays to perform query processing.
The first step for query execution in \system is to execute the \texttt{Scan}, \texttt{GroupBy}, and \texttt{Aggregation} operators in one RT job. The three core operators are adjacent in a query plan and, thus, can be executed together in a single job.
Please note that even if a query only has one or two operators, it can still be successfully executed when the BVH contains all the needed attributes.
\system determines the corresponding query area in the space based on \texttt{Scan} and launches a set of rays to intersect all primitives in the area. By only accessing data records in the query area defined by \texttt{Scan}, there is a significant reduction in the amount of accessed data for a query. Based on the coordinate of a primitive, \system can obtain its data attributes for \texttt{GroupBy} and \texttt{Aggregation} and then perform the corresponding operations.
For each data record, the data attributes involved in the three operators are stored together as the coordinate, which can be retrieved by only one memory access. It dramatically reduces the number of random accesses to the device memory.
After RT cores complete execution, \system executes the following operators, if any, in the CUDA cores, e.g., \texttt{OrderBy} and \texttt{Having}. In this way, \system is capable of handling a broad class of database queries.

Overall, \system effectively maps the query execution as an RT job, where the involved operators are executed simultaneously as one RT job with reduced data accesses and optimized data access pattern. With the acceleration from the ray tracing cores, \system is expected to bring a significant performance improvement.

\subsection{The Mechanism  of \system: An Example}

In this subsection, we demonstrate how a database table is mapped to the 3D space. Suppose there is a table named \(Score\), as shown in Figure~\ref{fig:query}a, that stores information about students' scores. Each row of the table corresponds to a student, and the table has many attributes, among which the \(Math\), \(English\), \(Class\), and \(Grade\) are used to store students' math scores, English scores, classes, and grades, respectively. 
We use query A (Section ~\ref{sec:bam:aeq}) to show how a query is mapped to rays for intersections. The query obtains the average math score of students in each class who have passed English.

\noindent\textbf{Mapping Data to Primitives:}
To perform queries with RT cores, \system first maps data records as primitives.
Based on the idea of representing data attributes involved in \texttt{Aggregation}, \texttt{GroupBy}, and \texttt{Scan} by the coordinates in three-dimensional space, \system makes the X-axis represent the data attributes involved in \texttt{Aggregation}, the Y-axis represent the data attributes involved in \texttt{GroupBy}, and the Z-axis represents the data attribute involved in \texttt{Scan}. In this way, each data record in the table corresponds to a point in space. 
Here, we assume the table is stored in a row-major format, or the table should be joined to get the corresponding coordinates for each axis.
For example, row 0 of the table corresponds to \((72, 7, 100)\) in space. Then, using the point as its vertex, \system creates a right triangle as the primitive. The reason why we choose triangles as primitives is that only the ray-triangle intersection test is hardware-supported by RT cores, while the intersection tests for other types of primitives are software-based and offloaded to CUDA cores. Therefore, the use of triangles allows \system to enhance performance by exploiting hardware acceleration from RT cores. Specifically, if the coordinate of a data record is \((a, b, c)\), then the three vertex coordinates of the right triangle we create are \((a, b, c)\), \((a+S_x, b, c)\), and \((a, b+S_y, c)\), where \(S_x\) and \(S_y\) are the two leg lengths of the right triangle. Therefore, the projections of a primitive on the XZ-plane and YZ-plane are a line segment $S_x$ and a line segment $S_y$, respectively. In this case, each data record in the table is mapped to a triangle in three-dimensional space, as shown in Figure~\ref{fig:query}a. Since the coordinates for all the primitives can be derived from the data attributes in the table, the BVH can be built in advance before query execution.

\begin{algorithm}[t]
\caption{Pseudo-code of Any Hit Shader}
\label{alg:anyhit}
\renewcommand{\algorithmicrequire}{\textbf{Input:}}
\renewcommand{\algorithmicensure}{\textbf{Output:}}
\begin{algorithmic}[1]
\REQUIRE flag bit array $V_{flag}$, result arrays $V_{sum}, V_{count}$
\ENSURE result arrays $V_{sum}, V_{count}$
\STATE $\text{primIdx} \gets \text{get\_prim\_index()}$
\STATE $[a, b, c] \gets \text{get\_prim\_right\_vertex\_coord(primIdx)}$
\STATE $\text{flag} \gets \text{atomic\_bit\_exch}(V_{flag}[primIdx], 1)$
\IF{$\text{flag} = 0$}
\STATE $\text{atomic\_add}(V_{sum}[b], a)$
\STATE $\text{atomic\_add}(V_{count}[b], 1)$
\ENDIF
\end{algorithmic}
\end{algorithm}

\noindent\textbf{Mapping A Query to Rays:}
With the built BVH, \system maps the query into a set of rays. For query \(A\), \system maps it to a set of parallel rays starting from the \(Z=60\) plane to the \(Z=100\) plane, along the positive direction of the Z-axis, as shown in Figure~\ref{fig:query}b. The rays launched should be dense enough to intersect all triangles in the region. Rays are launched as a two-dimensional array from the view of the XY-plane (\(Z=60\) plane), which have an interval of \(\frac{1}{2} S_x\) along the X-axis and an interval of \(\frac{1}{2}S_y\) along the Y-axis. Recall that when mapping data to primitives, the two legs of right triangles have lengths \(S_x\) and \(S_y\), respectively. The design guarantees that a triangle can be at least intersected by one ray. As shown in Figure~\ref{fig:query}c, primitives may intersect one ray (triangle A) or three rays (triangle B), and in the limiting case primitives intersect at most three rays (triangle C)\footnote{\url{https://forums.developer.nvidia.com/t/what-is-the-limiting-case-of-ray-triangle-intersection/309730/2}}. If the interval grows, there may be triangles that fail to intersect any ray.  In turn, if the interval gets smaller, it increases the probability that a triangle is intersected by more than one ray, which degrades the performance. 
Rays entirely cover the query area \(60 \leq Z\leq 100\), and triangles in the query area are bound to intersect rays, while triangles not in the query area are bound not to intersect any ray. 

For students who pass the English examination, their triangles are in the query area \(60 \leq Z\leq 100\). Thus, the set of triangles intersecting a ray is the set of students that satisfies the predicate of \texttt{Scan}. For each triangle that is intersected by a ray, the Y-coordinate of its right-angle vertex is used to find the group to which it belongs, while the X-coordinate is read to compute the aggregate function, respectively. The aggregate function is \textit{AVG} in Query A, and we maintain two arrays in the Any Hit Shader function, whose pseudo-code is shown in Algorithm~\ref{alg:anyhit}. The \textit{sum} array is to store the sum of \(Math\) for all students in each group, and the \textit{count} array is to store the number of students in each group. Indexing by the Y-axis coordinate, \system appends the X-coordinate to the corresponding element of the \textit{sum} array and increments the corresponding element of the \textit{count} array by 1. The Any Hit Shader will be called each time a ray finds an intersection with a triangle. 
Therefore, the flag bit array \(V_{flag}\) (line 3) is used to ensure that triangles are not double-counted, and atomic operation \(atomic\_add\) (lines 5 and 6) is used to avoid synchronization issues when tracing multiple rays in parallel. After the BVH traversal is complete, the sum of the \(Math\) (\(V_{sum}\)) is divided by the number of students (\(V_{count}\)) to obtain the average score of \(Math\) in each group. The approach can be adopted for executing other aggregate functions whose procedure is similar to that of \texttt{AVG}, including \texttt{SUM}, \texttt{COUNT}, \texttt{MAX}, and \texttt{MIN}.

\begin{figure*}[htbp]
    \centering
    \begin{minipage}[t]{0.6\textwidth}
        \centering
        \includegraphics[width=\textwidth]{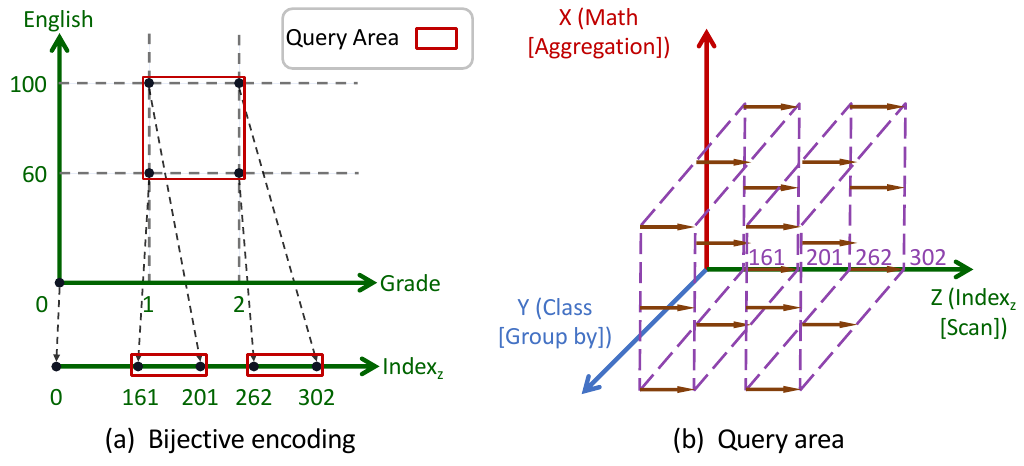}
        \caption{Processing for conjunctive predicates}
        \label{fig:conjuctivepredicate}
    \end{minipage}
    \hfill
    \raisebox{25mm}{
    \begin{minipage}[t]{0.35\textwidth}
        \begin{minipage}[t]{\textwidth}
            \centering
            \captionsetup{skip=2pt} 
            \includegraphics[width=\textwidth]{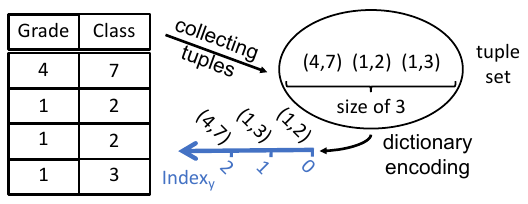}
            \caption{Encoding for GroupBy}
            \label{fig:group}
        \end{minipage}
        \begin{minipage}[t]{\textwidth}
            \centering
            \captionsetup{skip=2pt} 
            \includegraphics[width=\textwidth]{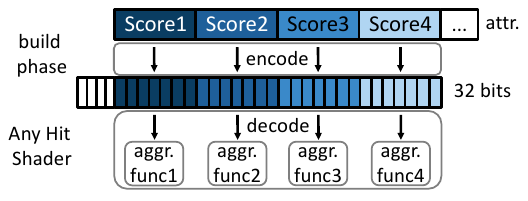}
            \caption{Encoding for Aggregation}
            \label{fig:aggr}
        \end{minipage}
    \end{minipage}
    }
\end{figure*}


\section{Encoding of Multiple Data Attributes}

\label{sec:ex}

\subsection{Motivation and Challenges}
Query \(A\) is a simple query where each operator involves only one data attribute. However, in real-world queries, it is common for an operator to involve multiple data attributes.
When mapping Query \(A\) in the space, a primitive directly uses the value in the corresponding attribute as the coordinate on an axis. For instance, \(English\) is used as the coordinate on the Z-axis.
However, when multiple data attributes are evaluated by one operator, the coordinate on one dimension needs to represent all data attributes involved.
For instance, with the Where clause \texttt{WHERE English $\ge$ 60 and Math $\ge$ 60}, both English score and Math score should be encoded in the Z-axis.

By studying the Star Schema Benchmark (SSB), it can be found that the number of data attributes involved in a single operator is generally small (no more than 4), and the ranges of most involved attributes are relatively narrow.
For \texttt{GroupBy}, there are two or three attributes involved in this operator in SSB, and the attributes that appear in a query are highly related.
For instance, attributes \texttt{d\_year} and \texttt{p\_brandl} are generally used in one \texttt{GroupBy} operator, while \texttt{c\_city}, \texttt{s\_city}, and \texttt{d\_year} are used as \texttt{GroupBy} attributes together in another class of queries.
The case also applies for \texttt{Aggregation} and \texttt{Scan}. Specifically, for \texttt{Aggregation}, only four attributes are used across all SSB queries.
Moreover, more than 80\% of the data attributes in SSB are within the range of 1000, which needs at most 10 bits for representation.
These properties offer an opportunity for efficient data compressing. 

Based on the above analysis, in \system, we propose to encode multiple data attributes as the coordinate on each axis.
However, since different operators have different functionalities, an appropriate encoding scheme needs to be chosen for each one. 
In this section, we study how to encode attributes for \texttt{Scan}, \texttt{GroupBy}, and \texttt{Aggregation}.

\subsection{Scan with Conjunctive Predicates}
\label{sec:ex:sub1}
The encoding scheme for \texttt{Scan} needs to maintain the relative order of the encoded data and specify the ray launching area to ensure correct execution.
A query generally contains multiple conjunctive predicates, like \(p_1 \wedge p_2 \wedge \cdots \wedge p_n\), and the attributes involved in the predicates can be encoded in the same attribute.

\textbf{Data Encoding:} \system adopts Bijection Encoding to encode multiple attributes in \texttt{Scan} as a coordinate. It treats all data attributes involved in conjunctive predicates as n-tuples, where each n-tuple is uniquely mapped to a natural number, and each natural number uniquely corresponds to an n-tuple. Assume that conjunctive predicates involves $n$ attributes \(A_1,A_2,\cdots,A_n\), where \(A_i \in [0, k_i)\) and \(A_i \in \mathbb{Z}\) (\(i \in [0, n]\)), then the encoding rule is as follows:
\[
Index_z = \sum_{i=1}^{n} \left[ \left( \prod_{j=i+1}^{n} k_j \right) A_i \right]
\]
If the data type of an attribute is not an integer or the data range of an attribute is not of the form \([0, k)\), a dictionary encoding is used to convert it to the above form.
Taking the query \(B\) as an example:

\begin{verbatim}
B: SELECT AVG(Math)
   FROM Score
   WHERE Grade between 1 and 2 AND English >= 60
   GROUP BY Class;
\end{verbatim}

For simplicity, it is assumed that the attributes involved in \texttt{Scan} are already dictionary-encoded.
Query \(B\) contains conjunctive predicates as its filtering conditions, which not only require students to have a passing score in English but also require students to be from either the first or second grade. \system assigns each \((Grade, English)\) tuple with a natural number by the following rule:
\[Index_z = 101 \cdot Grade + English\]
Here, \(101\) represents the range of \(English\) (\(English \in [0, 101)\), and \(English \in \mathbb{Z}\)). Figure~\ref{fig:conjuctivepredicate}a depicts the above process. 

\textbf{Ray Launching after Encoding:} After Bijective Encoding, the Z-axis represents \(Index_z\) generated by Bijective Encoding as an alternative. As a result, the query area changes as well. In the example, the query area is split into two parts, \(161 \leq Index_z \leq 201\) and \(262 \leq Index_z \leq 302\), as shown in Figure~\ref{fig:conjuctivepredicate}a. Correspondingly, as shown in Figure~\ref{fig:conjuctivepredicate}b, \system launches parallel rays from the \(Z=161\) plane to the \(Z=201\) plane and from the \(Z=262\) plane to the \(Z=302\) plane along the positive direction of the Z-axis to cover the entire query area. 
To note that \system tends to place attributes with larger predicate ranges later when determining the order of the attributes involved in the encoding, which helps reduce the complexity of partitioning the query area. In the preceding example, the query area is split into only two parts. However, if \(Grade\) and \(English\) are encoded in the reverse order, the query area will be split into 41 parts. More querying regions may impact the overall performance due to the increased ray launching latency.

\subsection{GroupBy with Multiple Attributes}
\label{sec:ex:sub2}
For data attributes involved in \texttt{GroupBy}, \system treats the $n$ data attributes as n-tuples, where each distinct n-tuple represents a different group. 
Take query \(C\) as an example, which groups students according to two data attributes \(Grade\) and \(Class\).

\begin{verbatim}
C: SELECT AVG(Math)
   FROM Score
   WHERE English >= 60
   GROUP BY Grade, Class;
\end{verbatim}

For all distinct n-tuples appearing in the table, \system maps them to natural numbers \(Index_y\) via dictionary encoding. The correspondence between \(Index_y\) and n-tuples is maintained by a mapping table that allows efficient lookup. After dictionary encoding, each natural number in \(Index_y\) represents a group, which serves as the Y-coordinate. Taking query \(C\) as an example, table \(Score\) is searched to find all distinct \((Grade, Class)\) tuples. Figure~\ref{fig:group} illustrates the above process. For simplicity of presentation, there are only 4 data records in the table. In the table, there are three different \((Grade, Class)\) tuples \((4,7)\), \((1,2)\), and \((1,3)\), making up a group set of size \(K=3\). Then, \system performs dictionary encoding to the group set, producing a coordinate  \(Index_y\) ranging from 0 to \(K-1\). As shown in the figure, \((1,2)\) is assigned to 0, \((1,3)\) is assigned to 1, and \((4,7)\) is assigned to 2. This encoding scheme facilitates the implementation of aggregation because the Y-coordinate can be directly used as the index to the array that stores aggregate results for each group.

\subsection{Aggregation with Multiple Attributes}
\label{sec:ex:sub3}
For \texttt{Aggregation}, we choose to pack the bits of the attribute values in a 32-bit float coordinate so that they can be directly used for the aggregate function. For example, assume that there is a query containing two aggregate functions \texttt{AVG(Math)} and \texttt{SUM(English)}.
Considering that the data range of course score is \([0, 100]\), we store \(Math\) and \(English\) with 7 bits each in the X-coordinate.
In Any Hit Shader, we obtain \(Math\) and \(English\) by decoding the X-coordinate and computing the two aggregate functions separately. 
Given that the coordinates have 32 bits and 7 bits are sufficient to store a single score, the X-coordinate can hold up to 4 scores. Therefore, the encoding can support queries with up to 4 aggregate functions, as shown in Figure~\ref{fig:aggr}. 

For cases where an aggregate function contains multiple data attributes, e.g., \texttt{SUM(Math + English)}, the calculated result \textit{Math + English} is directly encoded in the coordinate if the result does not exceed the expression range of a float.
Multiple queries in SSB have such forms of aggregations. 
This scheme may help make further compressions.
In this example, the range of \textit{Math + English} is $[0, 200]$, which only needs 8 bits instead of 14 bits when being separately stored. 
\system adopts this optimization when possible to store more attributes in a coordinate.

\subsection{Supporting More Attributes}
\label{sec:ex:sub4}
OptiX currently only supports single-precision floating-point numbers as coordinates, which express integers in $[-2^{24}, 2^{24}]$ without the loss of precision. This creates a restriction on the range of data attributes. The range of most data attributes in SSB is usually narrow enough to ignore the restriction. 
When multiple attributes are encoded into a coordinate with the encoding mechanism in \system, the limited expression range in OptiX may pose a challenge for certain queries. For \texttt{Scan} and \texttt{Groupby}, the encoded coordinate is an integer and stored as a float, where the number of encoded attributes is limited by the range after encoding. For \texttt{Aggregation}, the 32 bits are partitioned into several parts to store the involved attributes, and the number is limited by the sum of their least number of required bits.
Therefore, for a given query, there can be a situation in which the range of the encoded data exceeds the expression of a 32-bit float.

In this case, we encode as many attributes as possible in a coordinate while storing the rest attributes in the GPU device memory. When a primitive is intersected, the Any Hit Shader needs to access the corresponding attributes with the primitive ID. This approach can support a broader class of queries for \system, while it may influence query performance as it requires extra memory accesses.
With the rapid development of hardware-accelerated ray tracing technology, OptiX is expected to support higher precision data types as coordinates in the future to address this problem. 


\section{Handling Other Operators \& Fewer Operators}

\noindent\textbf{Other Operators} In addition to the three operators in the RT stage, i.e., \texttt{Scan}, \texttt{GroupBy}, and \texttt{Aggregation}, there can be other operators like \texttt{OrderBy}, \texttt{Having}, and \texttt{Top} in a query. We implement these operators as CUDA kernels since they are hard to map as an efficient RT job. Since these operators are generally performed after the operators in the RT stage, this makes \system support a broader class of queries.

\system makes optimizations on query execution and avoids the real execution of \texttt{OrderBy} in certain queries.
Because \system adopts a dictionary encoding for multiple data attributes involved in \texttt{GroupBy}, leading to a default order of the groups.  
Therefore, 
if the sequence of attributes in \texttt{OrderBy} is a prefix of the sequence of attributes in \texttt{GroupBy},
the results from the RT stage are already sorted in the specified order.
In this case, \texttt{OrderBy} does not need to be executed.
After studying various workloads, we find that there is an inherent \textit{nature order} for attributes in \texttt{OrderBy}.
For instance, we find that \texttt{order by year, nation} is a common pattern in SSB, and \texttt{nation} is always placed after \texttt{year}. Therefore, when encoding data attributes of \texttt{GroupBy}, we encode attributes in the \textit{nature order}, and it would not influence the performance of the RT job.

\noindent\textbf{Fewer Operators} In addition, there may be cases where \texttt{Scan}, \texttt{GroupBy}, and \texttt{Aggregation} are not all present in a query. For such a query, \system transforms it into an equivalent query with the pre-built BVH. For example, a query lacking \texttt{Scan} is considered to have a predicate with 100\% selectivity, where rays are launched to cover the entire space. When a query lacks \texttt{GroupBy}, it is considered to have only one group for all data records, and thus, only one variable is used for the aggregate function.


\section{Experimental Analysis}

\label{sec:ea}

\subsection{Experiment Setup}
\noindent\textbf{Hardware and Software} We run experiments on a machine equipped with two Intel(R) Xeon(R) Silver 4316 CPU @ 2.30GHz, 128GB DDR4 DRAM, and an NVIDIA GeForce RTX 4090 with 128 RT cores, 16384 CUDA cores, and 24GB VRAM. The operating system is 64-bit Ubuntu Server 20.04 with Linux Kernel 5.4.0-42-generic. The GPU programming interface uses CUDA 10.1 and OptiX 7.1. 

In Section 6.3, we also compare the performance between \system and Crystal on NVIDIA TITAN X (PASCAL), where Optix 5.1 is used for programming. The GPU was launched in 2015 and does not have RT cores. Therefore, ray tracing jobs are only executed on CUDA cores. The experiment aims to analyze the performance benefits from RT core acceleration.

\noindent\textbf{Workloads} Throughout the experiments, we adopt the Star Schema Benchmark (SSB)~\cite{SSB}, which has been widely used in various data analytics studies~\cite{ssbexample1,ssbexample2,ssbexample3,ssbexample4,Crystal,ssbexample5}. SSB provides a simplified star schema data based on the TPC-H benchmark~\cite{TPC-H}. There are a total of 13 queries in the benchmark, divided into 4 query flights. In our experiments, we run the SSB flat with a scale factor of 1, 10, and 20 to evaluate the performance with different data set sizes. When the scale factor equals 20, it will generate a flat table with 120 million tuples. 

\noindent\textbf{Baseline}
We compare \system with Crystal~\cite{Crystal}. Crystal is a recently proposed state-of-the-art GPU database system that delivers superior query execution performance compared to other systems. It currently supports only queries from the Star Schema Benchmark (SSB). 
Because \system joins columns when building the BVH, it would be unfair for the competitors to perform \texttt{Join} in performance comparison. Therefore, we adopt SSB flat, which flattens SSB into a wide table model. SSB flat is widely used in the industry to test the performance of the query engine~\cite{ClickHouse}.
In the evaluation, we emulate the implementation of Crystal on SSB queries and build a version that supports SSB flat. Besides, since Cyrstal did not implement the \texttt{OrderBy} operator, we removed it from SSB queries for a fair comparison.

\noindent\textbf{Encoding}
In SSB, we adopt different encoding schemes for the \texttt{Aggregation} operator.
Three queries in Flight Q4 have the aggregate function: \texttt{sum(lo\_revenue - lo\_supplycost)}, and we adopt the encoding optimization in Section~\ref{sec:ex:sub3} to make further compression.
The aggregate function of queries in Flight Q1 is \texttt{sum(lo\_extendedprice * lo\_discount)}. However, the range of \(lo\_extendedprice * lo\_discount\) is too large to be precisely represented as an integer by a 32-bit float. Therefore, in experiments, we use the approach in Section~\ref{sec:ex:sub4} to handle this situation. In evaluating Flight Q1, the X-coordinate only represents \(lo\_extendedprice\) while \(lo\_discount\) is stored in the GPU device memory.

\subsection{Query Execution Performance}
\label{sec:ea:sub1}
\begin{figure*}[htbp]
\begin{minipage}[t]{0.75\textwidth}
    \centering
    \begin{subfigure}[t]{0.32\textwidth}
        \centering
        \includegraphics[width=\textwidth]{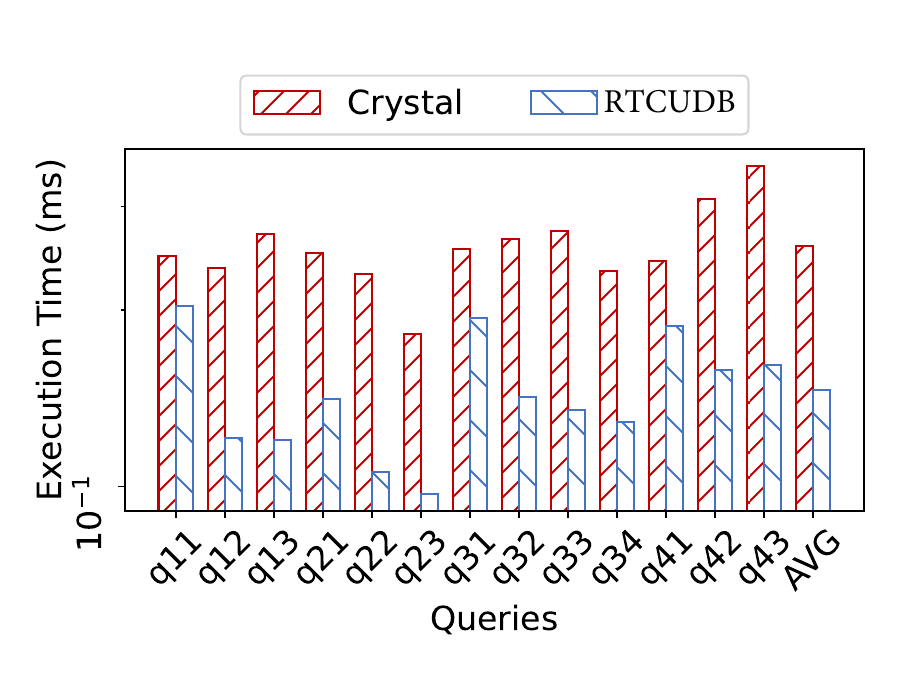}
        \caption{SF = 1}
        \label{fig:ExperimentA:sub1}
    \end{subfigure}
    \begin{subfigure}[t]{0.32\textwidth}
        \centering
        \includegraphics[width=\textwidth]{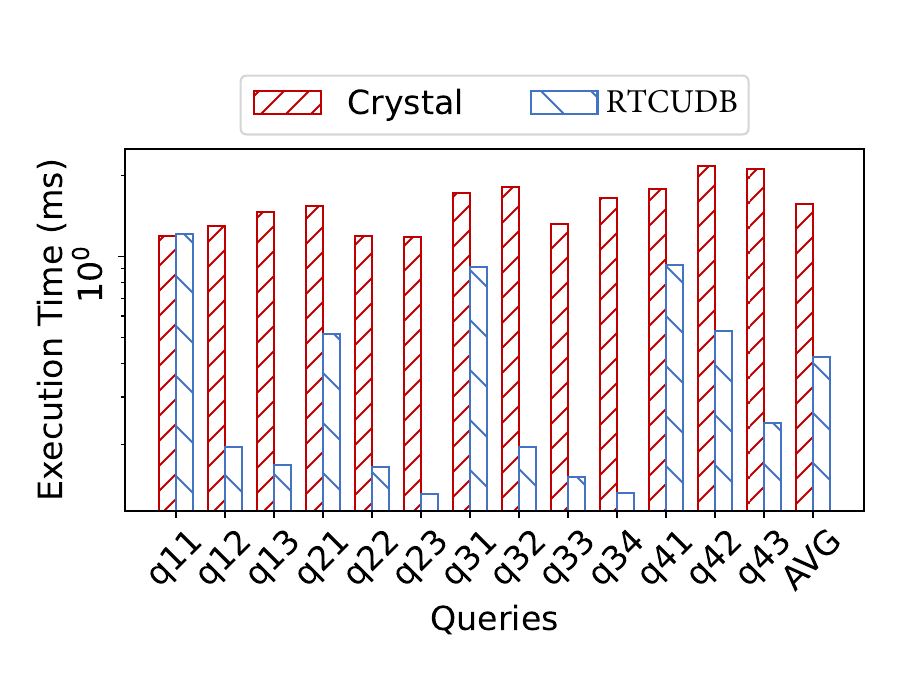}
        \caption{SF = 10}
        \label{fig:ExperimentA:sub2}
    \end{subfigure}
    \begin{subfigure}[t]{0.32\textwidth}
        \centering
        \includegraphics[width=\textwidth]{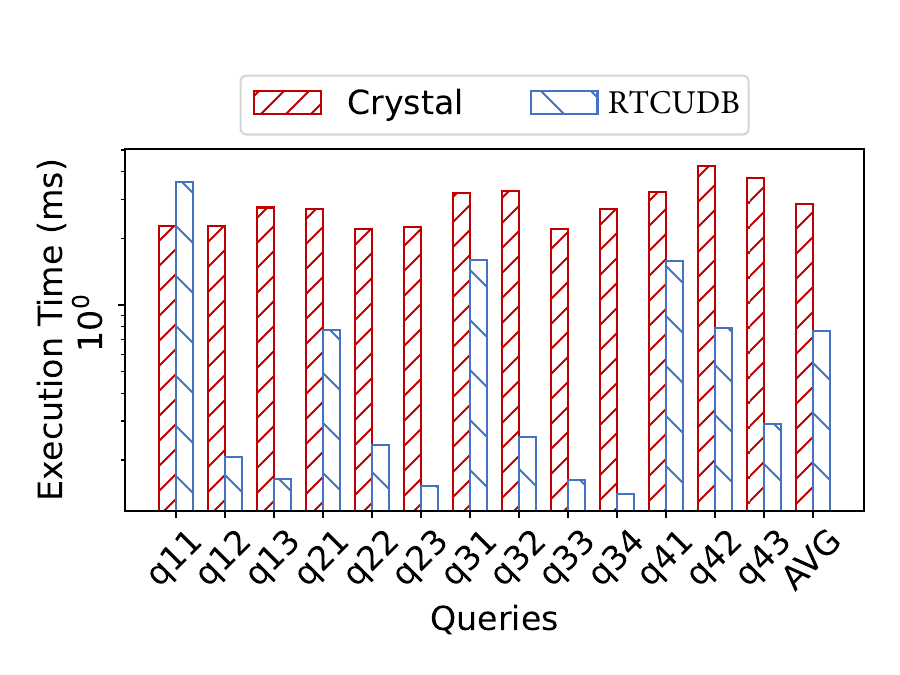} 
        \caption{SF = 20}
        \label{fig:ExperimentA:sub3}
    \end{subfigure}
    \caption{Query execution time of \system and Crystal}
    \label{fig:ExperimentA}
\end{minipage}
\begin{minipage}[t]{0.24\textwidth}
    \centering
    \includegraphics[width=\textwidth]{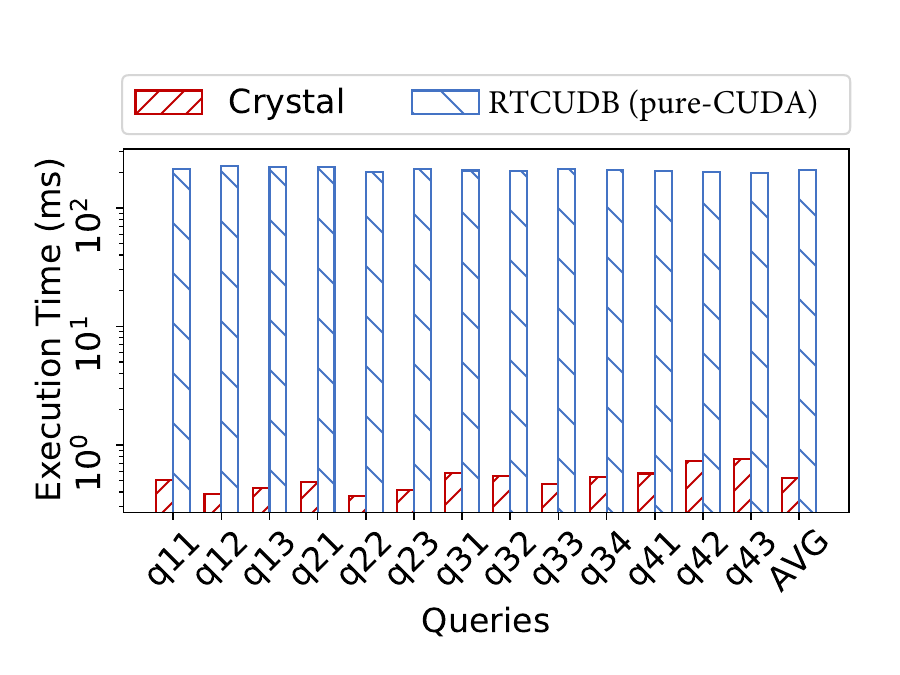}
    \caption{Performance comparison of \system (pure CUDA) and Crystal}
    \label{fig:ExperimentH}
\end{minipage}
\end{figure*}

Figure~\ref{fig:ExperimentA} illustrates the performance comparison between \system and Crystal. 
\system shows excellent performance on SSB flat. 
At SF=1, \system is faster than Crystal on all queries, on average, by 82.08\%.
At SF=10, \system is faster than Crystal on 12 out of 13 queries and \(5.4\times\) faster on average.
At SF=20, the situation is similar to that at SF=10. \system is faster than Crystal on 12 out of 13 queries, at least \(1.0\times\) faster and at most \(18.3\times\) faster. Over the entire SSB flat, \system is on average \(8.5\times\) faster than Crystal. 
It can be seen that \system maintains its performance advantage over Crystal in all SF cases. Considering that Crystal is by far the state-of-the-art GPU database system delivering superior query execution performance compared to other systems, the performance improvement is reasonably satisfactory. 

\begin{table}[t]
    \centering
    \setlength{\tabcolsep}{4pt} 
    \begin{tabular}{l|c c c c c c c }
        \toprule
        \textbf{Query} & q11 & q12 & q13 &  & q21 & q22 & q23 \\
        \midrule
        \textbf{Sel.(\%)} & 1.99 & 0.07 & 0.02 &  & 0.80 & 0.16 & 0.02 \\
        \toprule
        \textbf{Query} & q31 & q32 & q33 & q34 & q41 & q42 & q43 \\
        \midrule
        \textbf{Sel.(\%)} & 3.67 & 0.14 & 5.76E-3 & 7.33E-5 & 1.59 & 0.38 & 0.04 \\
        \bottomrule
    \end{tabular}
    \caption{Selectivity of queries in SSB}
    \label{tab:queryselectivity}
\end{table}

Table~\ref{tab:queryselectivity} records the selectivity of each query in SSB flat. Referring to Figure~\ref{fig:ExperimentA}, it is observed that the query execution time of \system has a certain positive correlation with the selectivity of the query. The selectivity of q11, q31, and q41 is significantly higher than the other queries, and their execution time is also longer. The fundamental reason for the positive correlation between query execution time and query selectivity under certain conditions is that a lower selectivity implies a smaller number of data records in the query area, that is, a smaller amount of data to be accessed. 
Among the above queries, although q11 does not have the highest selectivity (1.99\%, the maximum selectivity is 3.67\% of q31), its execution time is particularly long, and even the only query for which \system has a longer execution time than Crystal. The reason is the particularity of Flight q1. The queries in Flight q1 do not include the \texttt{GroupBy} operator. In our implementation, we treat such queries as if all data records belong to the same unique group. 
Recall that atomic operations are used in Any Hit Shader to avoid synchronization issues, but they also limit parallelism, which affects performance. In the presence of only one group, all atomic operations target the same scalar value, further preventing the parallelism advantage of RT cores from being fully exploited and slowing down the execution of q11. This is also confirmed by the shift of the execution time of q11 from SF=1 to SF=20: when the dataset is small (SF=1), there are also fewer atomic operations, and the restriction of parallelism is not obvious. Therefore, \system is faster than Crystal.  When the dataset becomes larger (SF=20), the atomic operations increase accordingly, and the disadvantage of limited parallelism is further highlighted. At this time, \system is inferior to Crystal.

Another observation is that the performance advantage of \system over Crystal becomes more and more pronounced as the dataset gets larger. \system is only 82.08\% faster than Crystal on average at SF=1, while it is \(8.5\times\) faster than Crystal on average at SF=20. To explain this phenomenon, we must first analyze the composition of the query execution time. The query execution time of \system can be thought of as roughly consisting of two parts. The first part is the ray generation time, which is related to the number of rays and remains relatively stable when the dataset size changes. The second part is the BVH traversal time, which is affected by the size of the dataset. In contrast, Crystal's execution time has no such fragmentation and is affected by the size of the dataset. When the dataset is small (SF=1), the performance advantage of \system is relatively small due to the existence of relatively fixed ray generation overhead. When the dataset is large (SF=20), the proportion of this part to the total execution time becomes irrelevant, and the performance advantage of \system is adequately highlighted.

\subsection{The Role Played by RT Cores in \system}
In order to justify the importance of RT cores for \system, we implement a pure-CUDA version of \system where the entire ray-tracing process is computed by CUDA cores on the GPU. To achieve this, we switch to OptiX 5.1, which uses CUDA cores to compute BVH traversal and ray-triangle intersection tests. Since OptiX 5.1 is an old version, it does not support NVIDIA RTX 4090. Therefore, the experiments in this subsection are performed on an NVIDIA TITAN X (Pascal). Limited by the device memory size of TITAN X, we conduct experiments with a scale factor of 1. The performance results are shown in Figure~\ref{fig:ExperimentH}.

On the 13 queries of SSB, \system (pure-CUDA) is \(258\times\)-\(588\times\) slower than Crystal, and \(423\times\) slower on average. Recall that \system is faster than Crystal on all queries with the same scale factor as in Section~\ref{sec:ea:sub1}. The two experimental results are in sharp contrast. This strongly confirms the crucial role played by RT cores in the effectiveness of \system. With the application of hardware-accelerated ray tracing technology, the BVH traversal and ray-triangle intersection tests that originally needed to be computed by CUDA cores during the ray tracing process are offloaded to the RT core, which is specialized hardware designed for this purpose, freeing CUDA cores from thousands of instructions per ray, which could be an enormous amount of instructions for an entire ray tracing process. The presence of RT cores considerably accelerates the ray-tracing process and makes the ray-tracing-based database possible.

\subsection{GPU Memory Bandwidth Occupancy}
\begin{figure*}[htbp]
    \centering
    \begin{subfigure}[t]{0.33\textwidth}
        \centering
        \includegraphics[width=\textwidth]{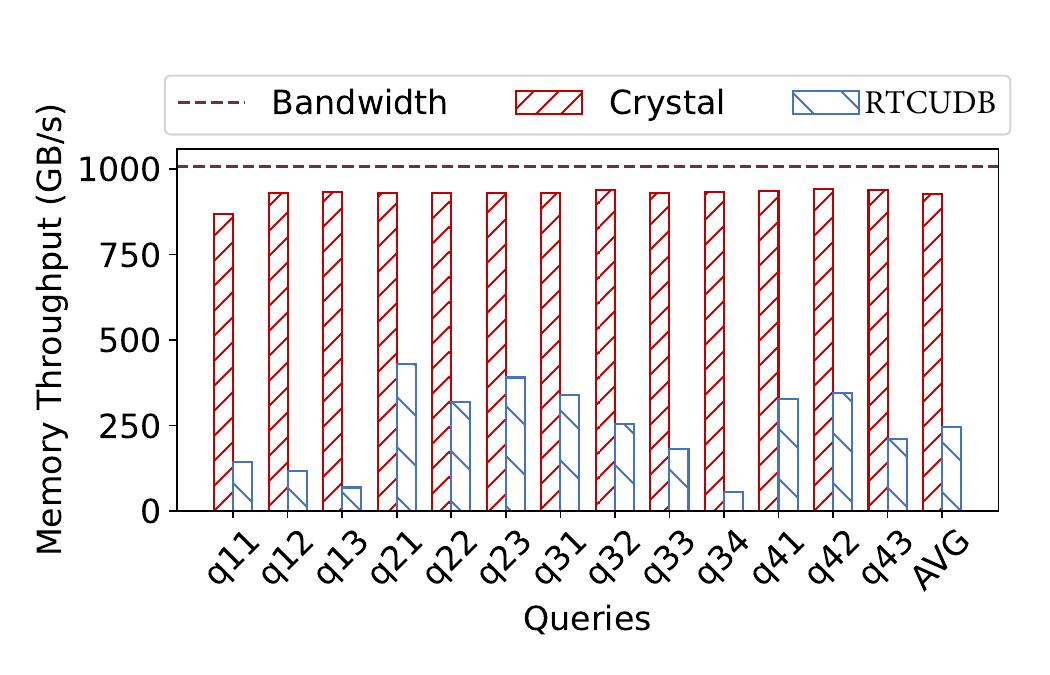}
        \caption{SF = 1}
        \label{fig:ExperimentC:sub1}
    \end{subfigure}
    \begin{subfigure}[t]{0.33\textwidth}
        \centering
        \includegraphics[width=\textwidth]{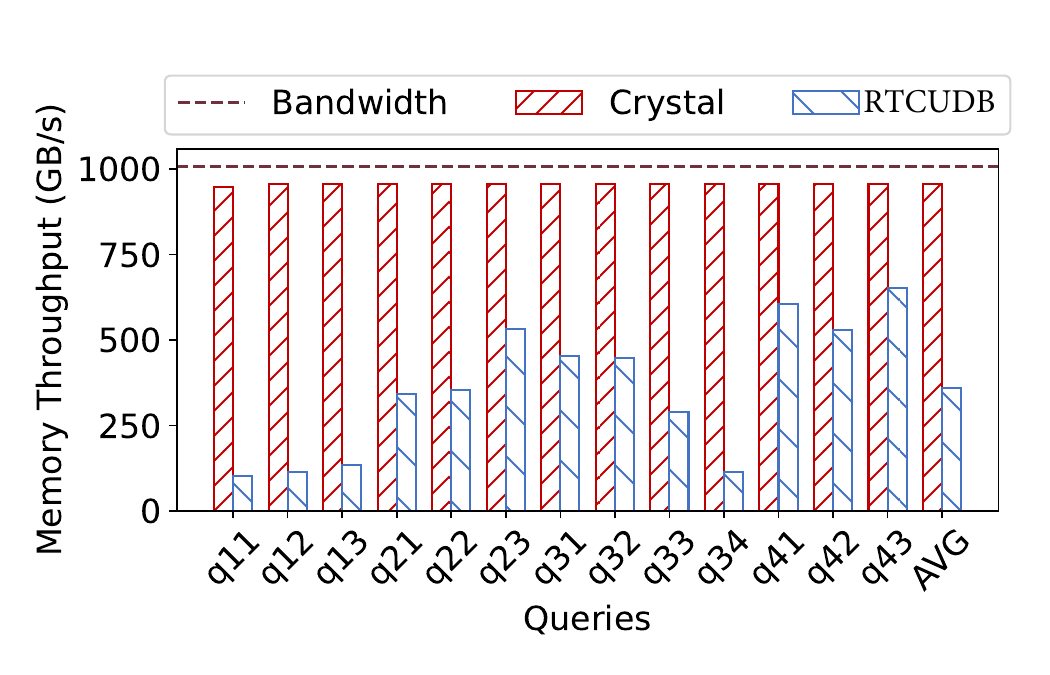}
        \caption{SF = 10}
        \label{fig:ExperimentC:sub2}
    \end{subfigure}
    \begin{subfigure}[t]{0.33\textwidth}
        \centering
        \includegraphics[width=\textwidth]{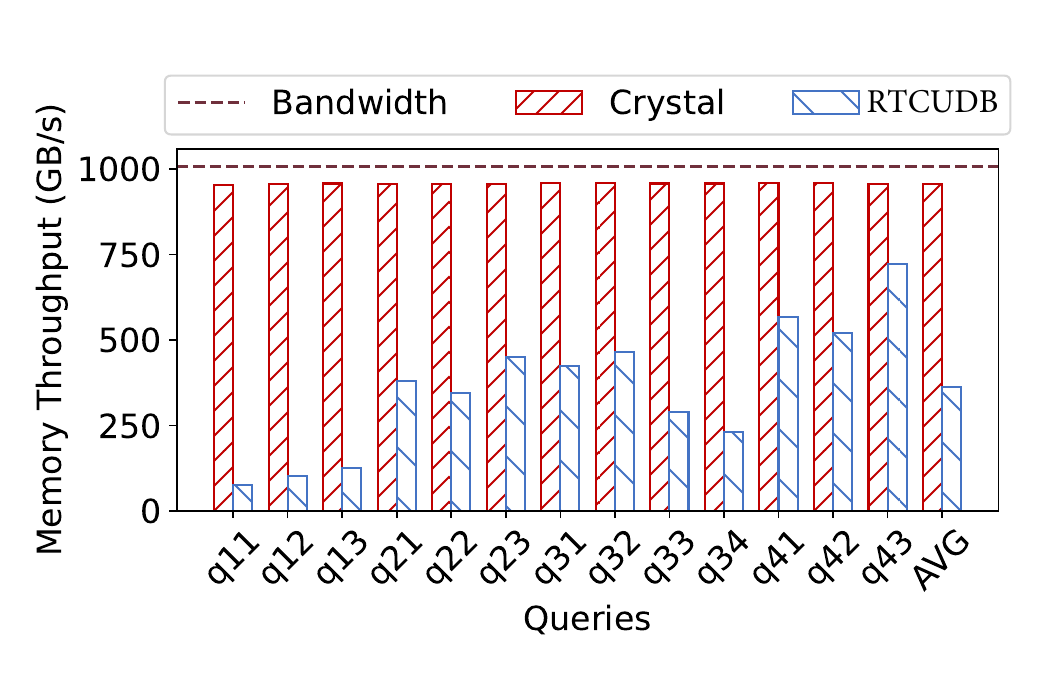} 
        \caption{SF = 20}
        \label{fig:ExperimentC:sub3}
    \end{subfigure}
    \caption{Query memory throughput of \system and Crystal}
    \label{fig:ExperimentC}
\end{figure*}

Figure~\ref{fig:ExperimentC} presents a comparison of the memory throughput of Crystal and \system on 13 queries of SSB flat. 
At SF=20, Crystal achieves a memory throughput of 97.11\% to 97.51\% of the memory bandwidth, with an average of 97.41\%. 
The situation is rather similar in other SF cases.
It can be argued that Crystal saturates the memory bandwidth. In contrast, 
at SF=20, \system's memory throughput is only 7.72\% to 73.53\% of the memory bandwidth, with an average of 36.74\%. 
This ratio is even smaller in other SF cases.
It can be seen that the memory bandwidth occupied by \system is considerably smaller than that of Crystal in all SF cases. This is due to the fact that \system drastically reduces the amount of data that needs to be accessed and the number of random memory accesses, freeing its performance from the memory bandwidth constraints. Based on this, we identify that the limiting factor for the performance of \system is the computational power of the RT cores. 

It is worth noting that there is no clear positive correlation between query memory throughput and query selectivity, which seems to contradict the conclusions we obtain. While \system reduces the amount of data that needs to be accessed with the help of ray tracing, it also increases the overhead of BVH traversal and ray-triangle intersection tests. Since the BVH is stored in the device memory, BVH traversal and ray-triangle intersection tests also require access to the memory, and this partial memory access is closely related to the BVH structure. The BVH structures corresponding to different queries show great differences, and their effect on the memory throughput shows strong stochasticity, ultimately shaping the results shown in Figure~\ref{fig:ExperimentC}. Nevertheless, overall, the query memory throughput is still significantly degraded.



\subsection{Comparison between Ray Tracing for the Entire Query and for Individual Operators}
In Section~\ref{sec:bam:aeq}, we explain why we choose to use ray tracing to accelerate the entire query rather than each operator. In the following, we conduct experiments to verify our conclusions. In view of the fact that there have been few studies on accelerating a single operator using ray tracing, which mainly focuses on the \texttt{Scan} operator (e.g., RTScan), we implement the \texttt{Scan} operator of the query by RTScan and the other operators by \system.
RTScan transmits its execution result to \system in the form of a bit vector, where each bit records whether the corresponding data record satisfies predicates. \system enforces the Z-coordinate to be 0 for all primitives. Accordingly, rays are launched at the \(Z=0\) plane as the query area. In this way, all primitives will be intersected by rays, and \system needs to use the bit vector to determine whether a data record satisfies the predicates and should continue to participate in the following computation. Since RTScan suffers from out-of-memory issues when SF=20, we down-regulate SF to 16 in this experiment.
The performance comparison with the case where \system implements the entire query is shown in Figure~\ref{fig:ExperimentD}. The performance of ray tracing for individual operators exhibits a noticeable decrease when compared to ray tracing for the entire query. On the 13 queries of SSB flat, ray tracing for the entire query is \(13.3\times\) - \(555.1\times\) faster than ray tracing for individual operators and \(194.9\times\) faster on average. The reason for this staggering performance gap is that ray tracing is used to accelerate \texttt{Scan} and other operators separately, which means that \system (disable \texttt{Scan}) requires access to the entire dataset, taking away the benefit of ray tracing. Overall, this result is sufficient to demonstrate the inefficiency of ray-tracing for individual operators and the necessity of choosing ray-tracing for the entire query. 

\subsection{Comparison between Encoding and Splitting}
In Section~\ref{sec:ex}, faced with the case where a single \texttt{Scan} and \texttt{GroupBy} operator involves multiple data attributes, we give a solution that is encoding. However, there is a more intuitive way, which is to split the data attributes involved in the operator into two parts.  The first part contains only one data attribute, which is still used for \system. The data attributes in the other part are stored in device memory in the form of arrays that can be accessed according to primitive indices. We refer to this approach as splitting.

\begin{figure*}[htbp]
    \centering
    \begin{minipage}[t]{0.33\textwidth}
        \centering
        \includegraphics[width=\textwidth]{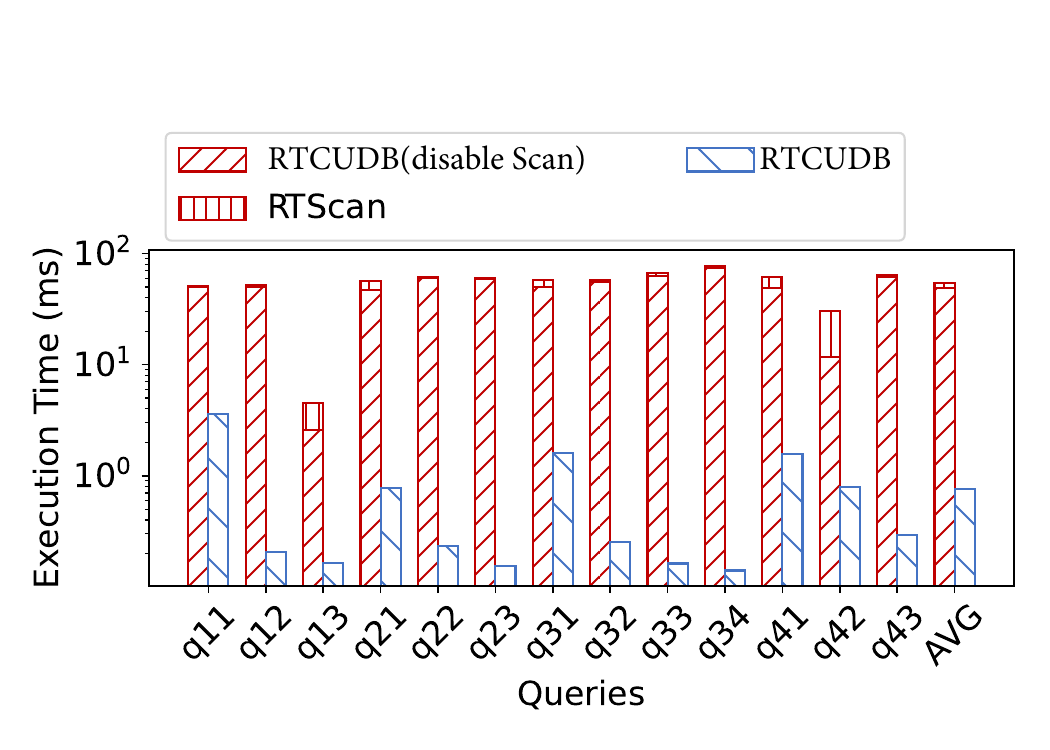}
        \caption{Performance comparison between \system and \system+RTScan}
        \label{fig:ExperimentD}
    \end{minipage}
    \hfill
    \begin{minipage}[t]{0.33\textwidth}
        \centering
        \includegraphics[width=\textwidth]{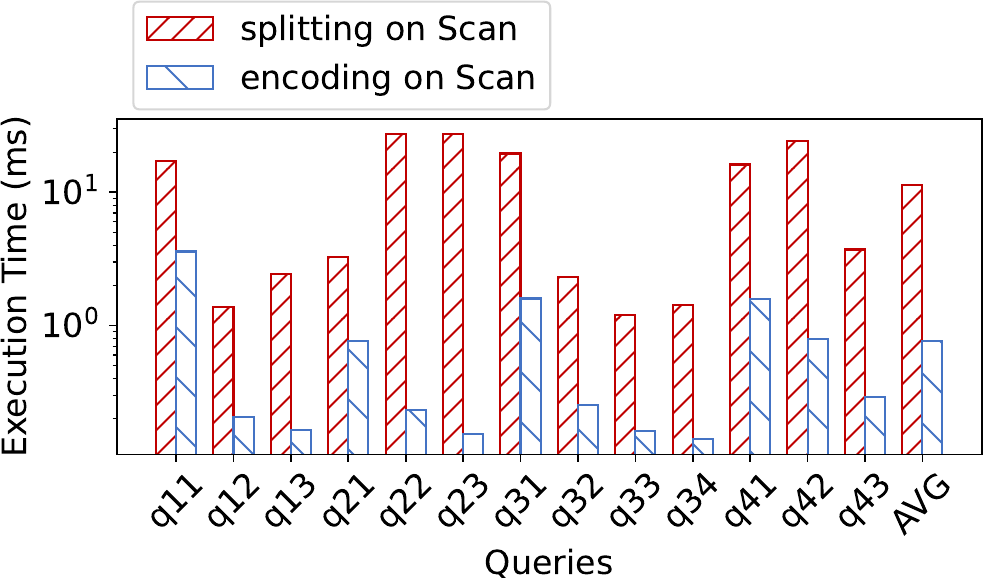}
        \caption{Performance improvement with encoding on Scan}
        \label{fig:ExperimentE}
    \end{minipage}
    \hfill
    \begin{minipage}[t]{0.33\textwidth}
        \centering
        \includegraphics[width=\textwidth]{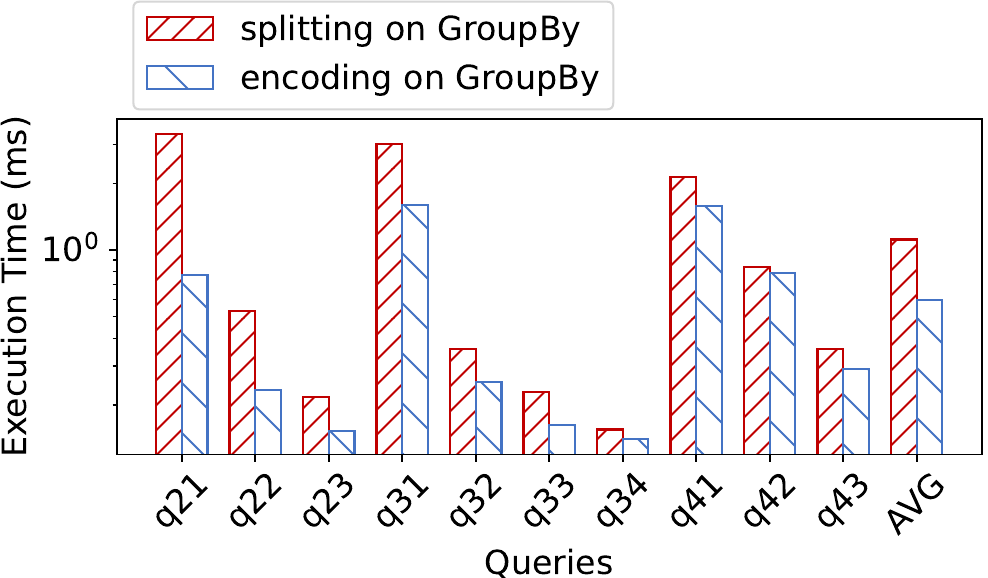}
        \caption{Performance improvement with encoding on GroupBy}
        \label{fig:ExperimentF}
    \end{minipage}
\end{figure*}

Figure~\ref{fig:ExperimentE} is a plot of the performance comparison between splitting and encoding on the \texttt{Scan} operator when SF=20. In the experiments, splitting on \texttt{Scan} uniformly selects the first data attribute involved in the \texttt{Scan} operator for \system and stores the remaining data attributes in the device memory. When RT cores detect a primitive that intersects a ray, the corresponding Any Hit Shader reads from the device memory the remaining data attributes of the corresponding primitive and determines whether they satisfy the conjunctive predicates contained in \texttt{Scan}. 
Experimental results show that encoding on \texttt{Scan} performs significantly better than splitting on \texttt{Scan}. on the 13 queries of SSB flat, encoding on \texttt{Scan} is \(3.3\times\) to \(176.9\times\) faster than splitting on \texttt{Scan}, with an average of \(31.1\times\). The reason for such a large performance difference is the difference in the amount of data that needs to be accessed. For example, if the conjunctive predicates are \(p_1 \wedge p_2\), \(p_1\) has a selectivity of 50\% and \(p_2\) has a selectivity of 20\%, then encoding on \texttt{Scan} will access 50\%\(\times\)20\%=10\% of the data records, and splitting on \texttt{Scan} will access 50\% of the data records. In this case, the amount of data they need to access differs by a factor of 5. Due to the different selectivity of each predicate in the conjunctive predicates, the performance difference between encoding on \texttt{Scan} and splitting on \texttt{Scan} on each query of SSB flat also shows a large difference. Overall, however, encoding on \texttt{Scan} outperforms splitting on \texttt{Scan} by a wide margin.

Figure~\ref{fig:ExperimentF} illustrates the performance comparison of splitting and encoding on the \texttt{GroupBy} operator when SF=20. 
Splitting on \texttt{GroupBy} is analogous to splitting on \texttt{Scan}.
Since the queries in Flight q1 do not involve the \texttt{GroupBy} operator, the experiment is performed only on the remaining 10 queries of SSB flat. It can be observed that the difference in performance between splitting on \texttt{GroupBy} and encoding on \texttt{GroupBy} is smaller, 74.61\% on average, compared to the difference between splitting on \texttt{Scan} and encoding on \texttt{Scan}, which is \(31.1\times\) on average. This is because splitting on \texttt{GroupBy} does not change the amount of data records accessed compared to encoding on \texttt{GroupBy} but only increases the overhead of accessing the remaining data attributes stored in the device memory. However, only the remaining data attributes of primitives that intersect rays will be accessed, so this overhead is not apparent in most SSB queries with generally low selectivity. 
Regardless, given the slight performance benefit, encoding on \texttt{GroupBy} remains the better option.

\section{Related Work}

\noindent\textbf{Applications of RT cores}
Since RT cores are originally designed to render physically correct reflections, refractions, shadows, and indirect lighting in real time~\cite{Turing}, there are numerous studies that leverage it to accelerate rendering workloads, including graphic rendering~\cite{graphic-rendering1,graphic-rendering2}, ambient occlusion~\cite{ambient-occlusion}, and simulation in physics~\cite{simulation-in-physics1,simulation-in-physics2}. In addition, there are also many studies that creatively use RT cores to accelerate non-rendering workloads such as data processing, including K-nearest neighbor search~\cite{RT-KNNS,RTNN}, database scan~\cite{RTScan,RTIndex}, range minimum queries~\cite{range-min-query}, point location search~\cite{point-location-search1,point-location-search2}, and rendering of unstructured meshes~\cite{rendering-of-unstructured-meshes}. 

\noindent\textbf{Accelerating queries with GPU}
GPUs offer strong parallelism and high-bandwidth memory, making them an attractive candidate for accelerating database queries. There are already various general-purpose GPU database systems available, including HeavyDB~\cite{HeavyDB}, BlazingSQL~\cite{BlazingSQL}, and TQP~\cite{TQP}. There are three types of computational cores on the GPU: CUDA cores, Tenser cores, and RT cores. CUDA cores are responsible for integer and floating-point operations, so most GPU database systems are based on CUDA cores. Among them, Crystal~\cite{Crystal} is the state-of-the-art CUDA-based GPU database that provides superior query execution performance compared to other systems. Tensor cores provide significant speedups to matrix operations. To leverage the computational power of Tensor cores, TCUDB~\cite{TCUDB} maps query operators to efficient matrix operators and implements a Tensor-based GPU database. RT cores accelerate Bounding Volume Hierarchy (BVH) traversal and ray-triangle intersection tests in ray tracing, and \system is the first study to make use of them to accelerate database queries, filling the gap in related directions.

\section{Conclusion}

In this paper, we propose \system, a query engine that efficiently maps database queries to ray tracing jobs, which effectively exploits RT cores for acceleration. 
Instead of implementing each operator independently like CUDA-based implementations, \system maps the core operators in a query as one RT job.
The approach brings several performance advantages, including accessing data with an optimized sequential access pattern, reducing the amount of data to be accessed, and exploiting the parallelism of RT cores. \system breaks the memory bandwidth restriction on query performance and significantly outperforms the state-of-the-art CUDA-based GPU and CPU query engines. 




\bibliographystyle{plain}
\bibliography{b}

\end{document}